\documentclass[12pt,preprint]{aastex}

\def\({\left(}
\def\){\right)}

\shorttitle{Power spectrum estimatation}
\shortauthors{Scodeller, Hansen}

\begin{document}

\title{Masking versus removing point sources in CMB data: the source corrected WMAP power spectrum from new extended catalogue}

\author{Sandro Scodeller}
\affil{Institute of Theoretical Astrophysics, University of Oslo,
P.O. Box 1029 Blindern, N-0315 Oslo, Norway}
\email{sandro.scodeller@astro.uio.no}

\author{Frode K. Hansen}
\affil{Institute of Theoretical Astrophysics, University of Oslo,
P.O. Box 1029 Blindern, N-0315 Oslo, Norway; \\Centre of Mathematics
for Applications, University of Oslo, P.O. Box 1053 Blindern, N-0316
Oslo }
\email{frodekh@astro.uio.no}

\begin{abstract}
In (Scodeller et al.) a new and extended point source catalogue obtained from the WMAP 7-year data was presented. It includes most of the sources included in the standard WMAP 7-year point source catalogues as well as a large number of new detections. Here we study the effects on the estimated CMB power spectrum when taking the newly detected point sources into consideration. We create point source masks for all the 2102 sources that we detected as well as a smaller one for the 665 sources detected in the Q, V and W bands. We also create WMAP7 maps with point sources subtracted in order to compare with the spectrum obtained with source masks. The extended point source masks and point source cleaned WMAP7 maps are made publicly available. Using the proper residual correction, we find that the CMB power spectrum obtained from the point source cleaned map without any source mask is fully consistent with the spectrum obtained from the masked map. We further find that the spectrum obtained masking all 2102 sources is consistent with the results obtained using the standard WMAP 7-year point source mask (KQ85y7). We also verify that the removal of point sources does not introduce any skewness.
\end{abstract}


\section{Introduction}
\label{sec:intro}

The study of the Cosmic Microwave Background (CMB) is a fundamental tool for understanding the universe we live in. A very important link from the study of the CMB to parameters describing our universe is the power spectrum $C_\ell$ of the anisotropies in the CMB (see for instance \cite{wmap7_power}). It is hence very important to accurately measure the spectrum $C_\ell$, which is obtained via measuring the temperature of the sky, which in turn has contaminating foregrounds. The main contaminants on larger scales are extended foregrounds (diffuse galactic emissions) while on small scales, extra-galactic point sources are the main contaminants (see for instance \cite{toff98}, \cite{DeZott99}, \cite{Hobs99}, \cite{DeZott05} or \cite{counts}). The most accurate (meaning high resolution and high signal-to-noise ratio) publicly available measurements of the CMB anisotropies we have up to today are from the Wilkinson Microwave Anisotropy Probe (WMAP) \citep{WMAP03}. Clearly, measuring the CMB and its anisotropies accurately also implies correcting accurately for the contaminants. In \cite{newdet}, we presented a new way to disentangle the point source signal from the CMB and other foregrounds obtaining a new and extended catalogue of extragalactic sources. The novelty of our approach was the use of needlet transforms (see \cite{need} or \cite{scodeller} and references therein) and internal templates (\cite{IT}) to enhance the signal-to-noise ratio of the point source signal. 

The point sources contribute to the spectrum mainly at large multipoles $\ell$ and correcting the power spectrum from the bias that they introduce is crucial for accurately estimating the cosmological parameters. This can be done in two ways: the standard approach is to mask the observed map with a point source mask and then compute the spectrum (see (\cite{wmap7_power})). Another approach is to remove the best model estimates of the point sources in the map and then compute the power spectrum, which to our knowledge has been considered in only a very few papers (see \cite{Tegma} or \cite{Bajko}). The advantage of the standard approach is that it has no dependence on whether the source is modelled correctly, but each source is masked with a rather large disc depending on the beam of the experiment and some data are lost. Another disadvantage with this approach is that analysis methods applying filters (for instance wavelets) to the map may contaminate an even larger part of the filtered map in the area of the point source holes (see for instance \cite{scodeller}). The alternative approach depends on modelling the source correctly, but less data are masked and filtered maps will not be contaminated by point source holes. In this paper we use both approaches in order to compare them. Even if we increased substantially the number of detected point sources, there are still unresolved point sources which contribute to estimate of the spectrum $C_\ell$. We use the approach first presented in \cite{unres} by the WMAP-team to correct for it. In short, this method consists in using the CMB power spectrum (containing the unresolved point sources contribution) obtained at different frequencies and the known frequency dependency of the flux densities of the point sources (see for instance \cite{wmap5} or \cite{Lanz11}) in order to make a fit of the amplitude of the unresolved point sources.

The scope of this paper is threefold:
\begin{enumerate}
\item To compare the effect of masking point sources versus removing best-fitting source models when estimating the CMB power spectrum. We do this by running a large number of simulations with point sources and compare the mean of the estimated spectra using each of the two procedures. Since the WMAP team used only the V and W channels to estimate the power spectrum and to save CPU time, we simulate only the V and W channels.
\item To compare the power spectrum in the WMAP 7-year data using the two different approaches above as well as to see the effect of masking/removing all the additional point sources detected in \cite{newdet} on the power spectrum.
\item To compare the masking and removing approach on the skewness of needlet coefficients.
\end{enumerate}

The outline of the paper is as follows: in \S \ref{sec:Meth} we present the methods both for creating simulations (\S\ref{sec:create_sims}) as well as the method to estimate the power spectrum  $C_\ell$ (\S\ref{sec:Method}). In \S \ref{sec:Results} we present the results both for simulations (\S\ref{sec:sim_res}) and WMAP 7-year data (\S\ref{sec:realdata}). In \S \ref{sec:skew} we present the results of the skewness analysis. Finally, in \S \ref{sec:concl} we present our conclusions.

\section{Method}

\label{sec:Meth}
\subsection{Creating the simulations}
\label{sec:create_sims}
To fulfill goals 1 and 3 above, we need to generate an ensemble of realistic simulations of CMB, noise and point sources for the V and W channels. To generate the point source simulations we draw the flux densities from a flux density distribution in the V-channel and transform to other channels drawing from a spectral index distribution. For the V-channel, we sample flux densities $S$ from the distribution for the differential number counts $\frac{dN}{dS}$ found in \citep{newdet},
\[\frac{dN}{dS}\propto S^{-2.14} \;\; \forall S>S_\mathrm{min},\]

We chose to sample flux densities in a range with lower bound $S_{min}=0.08$ Jy (in analogy to \cite{bias}), well below the detection limit, up to $6\langle\sigma_V\rangle\approx0.222$ Jy, where $\langle\sigma_V\rangle$ is the mean of the error on the estimated flux densities in the V channel. Above $6\langle\sigma_V\rangle$ we do not sample flux densities from the distribution for the differential number counts given above, but we use the actual flux densities estimated in the WMAP 7-year data. This choice is motivated by the fact that the point sources with high flux densities are well known in real data (and their number count is complete); if we were to simulate sources above $6\langle\sigma_V\rangle$, there being only a few (being at the tail of the distribution) the resulting simulated sky would look very different from the measured one.

We normalize the distribution function $dN/dS=nS^{-2.14}$ (giving the number of sources per Jy per steradian), where $n$ is the normalization constant. This constant is obtained from the number of detections of point sources in real data (V-band) which have flux densities above $\geq 6\langle\sigma_V\rangle$, which we found to be 103 detections  \citep{newdet},

\[
n\int_{6\times0.222 Jy}^{\infty}S^{-2.14}dS=103/(4\pi f_\mathrm{sky}),
\]
 where $f_{sky}$ is the kept sky-fraction.
The total number N of sources above $S_\mathrm{min}$  for the full sky is then determined by integrating
\[
N=n\int_{0.08 Jy}^{\infty}S^{-2.14}dS\times4\pi=3082
\]
Eventually this gives 2955 simulated flux densities being $<6\langle\sigma_V\rangle$. We then use the actual flux densities of the point sources being $>6\sigma_V$ in the V channel and the flux densities estimated in the V-channel of the sources which are $>6\sigma_W$ in channel W but $<6\sigma_V$ in the V-channel (which add up to 113 sources). We thus obtain a set of 3068 realistic values of flux densities of point sources at frequency V of which 2955 are drawn from the distribution of the differential number counts and 113 are flux densities of strong sources from real data.

In order to obtain realistic values of flux densities in the W-channel, we scale the flux densities $<6\langle\sigma_V\rangle$ in the V-channel with a spectral index distribution having mean $-0.28$ and a dispersion of 0.65. The mean value is chosen  half way from the mean spectral index obtained by \cite{Lanz11} (namely $-0.5$) and the WMAP-team \cite{wmap5} (namely $-0.09$), while for the dispersion we took the bigger one of the two, i.e. from \cite{Lanz11}.

We therefore sample 2955 values of spectral indices from a gaussian distribution with mean $-0.28$ and dispersion 0.65 and rescale the simulated flux densities for the V-channel with these spectral indices to simulated flux densities in the W-channel. Similarly to the approach in the V-channel we use the 113 actual flux densities from real data which are bigger then $>6\langle\sigma\rangle$ in one of the two channels.

Except for the 113 strong sources, for which we use amplitude and position estimated from real data, the position of the remaining sources are simulated randomly across the full sky. Since in \cite{newdet} we identify two point source as the same if their separation is less than $0.4^\circ$, we simulate all those sources which are stronger than the smallest detected flux density in V-band WMAP data at least $0.4^\circ$ from each other (but there is no such condition with respect to weaker sources).

In order to get a good estimate on the uncertainty of estimated power spectra we create 10000 realisations, for both the V and W channel, of CMB fluctuations (based on the WMAP best-fitting $C_\ell$, \cite{wmap7_power}), noise and then add the (same) pure point source map obtained as explained above.

\subsection{Influence of mask and point source removal on $C_\ell$: Method}
\label{sec:Method}

 The basic idea is to estimate and compare the power spectrum $C_\ell$ obtained by either removing or masking detected sources.  We will estimate the two autospectra VV and WW, the cross-spectrum VW as well as the optimal noise-weighted co-added combination of the autospectra. Our procedure will be applied to the full ensemble of simulated maps as well as to the actual WMAP7 data. Note that some steps are slightly different when using real data rather than simulations. When this is the case it will be shown in the step-by-step description in the following paragraph. The reason for these differences is the need to save CPU time.

Given that we have a set of simulated or actual maps for the V- and W-band, step-by-step our procedure is:
\begin{enumerate}
\item Use the detection algorithm presented in \citep{newdet} to detect point sources: 
\begin{description}
\item[Simulations:]
Sources are counted as detected in the V-band if
\begin{itemize}
\item they are found at $>5\sigma$ directly in V. 
\item they are found at $>5\sigma$ in W: In this case we search in V in a disc of radius $0.12^\circ$ around the position in W and if a source at $>3\sigma$ is found, it is counted as detected in V. This is different from the approach used in \cite{newdet} where we tested for $>5\sigma$ in the internal templates and then searched for $>3\sigma$ in the given band (note that the radius of the disc to search was defined from the error bars on the position of the source, this is what we have used to obtain $0.12^\circ$ in the present case). Since we only simulate the V and W channels, we only have one single internal template available with a small frequency difference making the internal template approach inefficient.
\end{itemize}
We use the same approach to define detected sources in the W-band.
\item[WMAP 7-year data:] We use the 665 point sources detected in the Q, V and W channel as presented in \cite{newdet}. We use both the point sources which were detected at $5\sigma$ in the channels and those detected using the internal template approach as described in the preceding paragraph. 
\end{description}

\item Construct the co-added map defined as the noise-weighted sum of the two frequency bands: \[M_{VW}=w_VM_V+w_WM_W,\] where $M_{VW}$ is the combine map, $M_i$ the map at channel $i$ and the weights $w_i$ are given by:\[w_i=\sum_p 1/\sigma_{pi}^2/\(\sum_p 1/\sigma_{pV}^2+\sum_p 1/\sigma_{pW}^2\)\] with $\sigma_{pi}$ being the rms in pixel $p$ of the instrumental noise in channel $i$ and the sum is over all pixels $p$.
\item Multiply the maps $M_V$, $M_W$ and the co-added map $M_{VW}$ with the KQ85y7 galactic mask (we would like to stress {\it galactic}, no point sources at all are masked in this step).
\item 
\begin{description}
\item[Simulations:] Remove the best-fitting models of the detected sources in each channel to create the maps $M_V^R$, $M_W^R$ and $M_{VW}^R$.
\item[WMAP 7-year data:] Remove the 665 detected point sources for which we have flux density estimates both from the band-averaged Q, V and W channels as well as from the 8 individual differencing assemblies (DAs) Q1 until W4. The different DA-maps are needed for better estimation of the unresolved point source amplitude. We also would like to point out that we only estimate the power spectrum in the V- and W-band, so in principle the Q-band is not needed. However, to estimate the contribution of unresolved sources in real data, we also need the Q-band to get smallest possible error bars (note that we need the Q-band only to obtain the contribution from unresolved sources, it is not used further). This is the reason why we also need to remove sources in the Q-band and Q-DAs.
\end{description}
\item 
\begin{description}
\item[Simulations:] Create a mask which masks the detected sources in each channel with radius of $0.6^\circ$\footnote{NB: in analogy to the WMAP approach, we use the same mask for the V and W channels, hence masking for instance a source in W, which is only detected in V. }. Then we multiply the 3 maps with point sources with the mask obtaining $M_{VW}^M$, $M_V^M$ and $M_W^M$. 
\item[WMAP 7-year data:] Create a mask which masks the 665 detected sources in each channel with radius of $0.6^\circ$. In addition, we also use different masking schemes, as will be described in detail in \S\ref{sec:realdata} and like in the preceding step we mask both the band-averaged channels and the DAs.
\end{description}
\item Estimate the power spectrum $C_\ell$ using the master algorithm \citep{master} with different channel-combinations obtaining: $C_\ell^{R,VW}$, $C_\ell^{R,VV}$, $C_\ell^{R,WW}$, $C_\ell^{M,VW}$, $C_\ell^{M,VV}$, $C_\ell^{M,WW}$. Estimate also the cross-power-spectra $C_\ell^{R,X}$ and $C_\ell^{M,X}$. Note that for multipoles $\ell>600$ we use inverse-noise weighting, like it was done by the WMAP-team in the 7-year analysis, see \cite{wmap7_power}.
\item Correct for a small bias in the power spectrum introduced by the residuals of the removed sources. We know that at the position of a removed source, there is a residual source with mean zero and standard deviation given by the error bar on the amplitude of the source. Furthermore, this residual sources have a slightly asymmetric form due to the uncertainty in the estimated position. 
\begin{description}
\item[Simulations:] Correct for this residual source bias in the power spectrum by taking the mean spectrum of 1000 source subtracted simulated maps where we subtract the pure CMB power spectrum. We use the same approach to check if there are residuals outside the point source holes when masking, but find that the correction in this case is negligible.
\item[WMAP 7-year data:] In the V and W channels we use the same procedure as for the simulations,  but the flux densities of the input sources used in the simulations of 1000 source subtracted maps for the bias estimation were all taken from actual values estimated in real data. As explained in step 4, we also need the Q-band for estimating the unresolved point source amplitude to higher precision. In order to save CPU time and avoid having to run full simulations for the Q-band to obtain the residual source correction, we run simplified simulations which we have verified give consistent values for the unresolved source contributions. This consists of pure source simulations where we subtract random model sources based on the standard deviations of the estimated amplitudes and positions. The only difference with the full approach used for the V- and W-band is that the correlation between estimated amplitude and the background CMB and noise is missing.
\end{description}
\item Correct for the contribution of the unresolved point sources contribution following the method of the WMAP team, first presented in \cite{unres}. Some simplifications were made with respect to the procedure of \cite{unres}. Namely:

  \begin{itemize}
\item We use the following power spectra of the CMB obtained in different bands:
\begin{description}
\item[Simulations:] the band-average (not the different DAs per band) spectra and the only three available spectra (VV, VxW and WW) are used, hence we are also using auto-spectra
  \item[WMAP 7-year data:] all 28 DA cross combinations (eg. Q1Q2,Q1V2,W3W4, etc.) are used (like in \cite{unres})
\end{description}
  \item  in the calculation of the correlation matrix $\Sigma^{\alpha\beta}_{\ell\ell}$, the WMAP team uses a multipole dependent sky fraction $f_{sky}(\ell)$ based on an effective noise power. Here instead, we use a slightly simplified procedure to obtain $f_{sky}(\ell)$ calibrated from 10000 simulations of CMB+noise. A bit more in detail, we compute the standard deviation of the estimated power spectrum $\Delta\hat{C}_\ell$ of the 10000 simulations. Then we obtain $f_{sky}(\ell)$ from:

    \begin{equation}
      f_{sky}(\ell)=\sqrt{\frac{2}{2\ell+1}}\frac{C_\ell+N_\ell}{\Delta\hat{C}_\ell},
    \end{equation}
    
    where $C_\ell$ is the best-fitting power spectrum from WMAP (input in the simulations) and $N_\ell$ is the noise power, computed with the appropriate mask.
  
    \item define the covariance matrix $\Sigma^{\alpha\beta}_{\ell\ell}$, where $\alpha$ and $\beta$ are channel combinations (for simulations: VV,VxW or WW, for WMAP 7-year data: the 28 DA cross combinations), with $f_{sky}(\ell)$ and $N_\ell$ obtained as described above.
    \item The uncertainty on the estimated amplitude $A_{ps}$ is given by:
 \begin{description}
\item[Simulations:] the standard deviation of $A_{ps}$ of the 9000 simulations (only 9000, because we used 1000 simulations for estimating the bias of the residuals).
 \item[WMAP 7-year data:] the optimal estimator as presented in \cite{unres}
\end{description}
 \end{itemize}
\end{enumerate}

 As a final remark on the method we would like to stress that when removing point sources one has to account for the bias introduced by the uncertainty on estimating the point source position and amplitude. The results in the following section show though that the two approaches are consistent, making it possible to estimate the power spectrum using only a Galaxy mask.

\section{Results}
\label{sec:Results}
\subsection{Simulations: Results}
\label{sec:sim_res}

\begin{figure}[!ht!p] 
  \centering
  \includegraphics[height=9.5 cm,width= 16 cm]{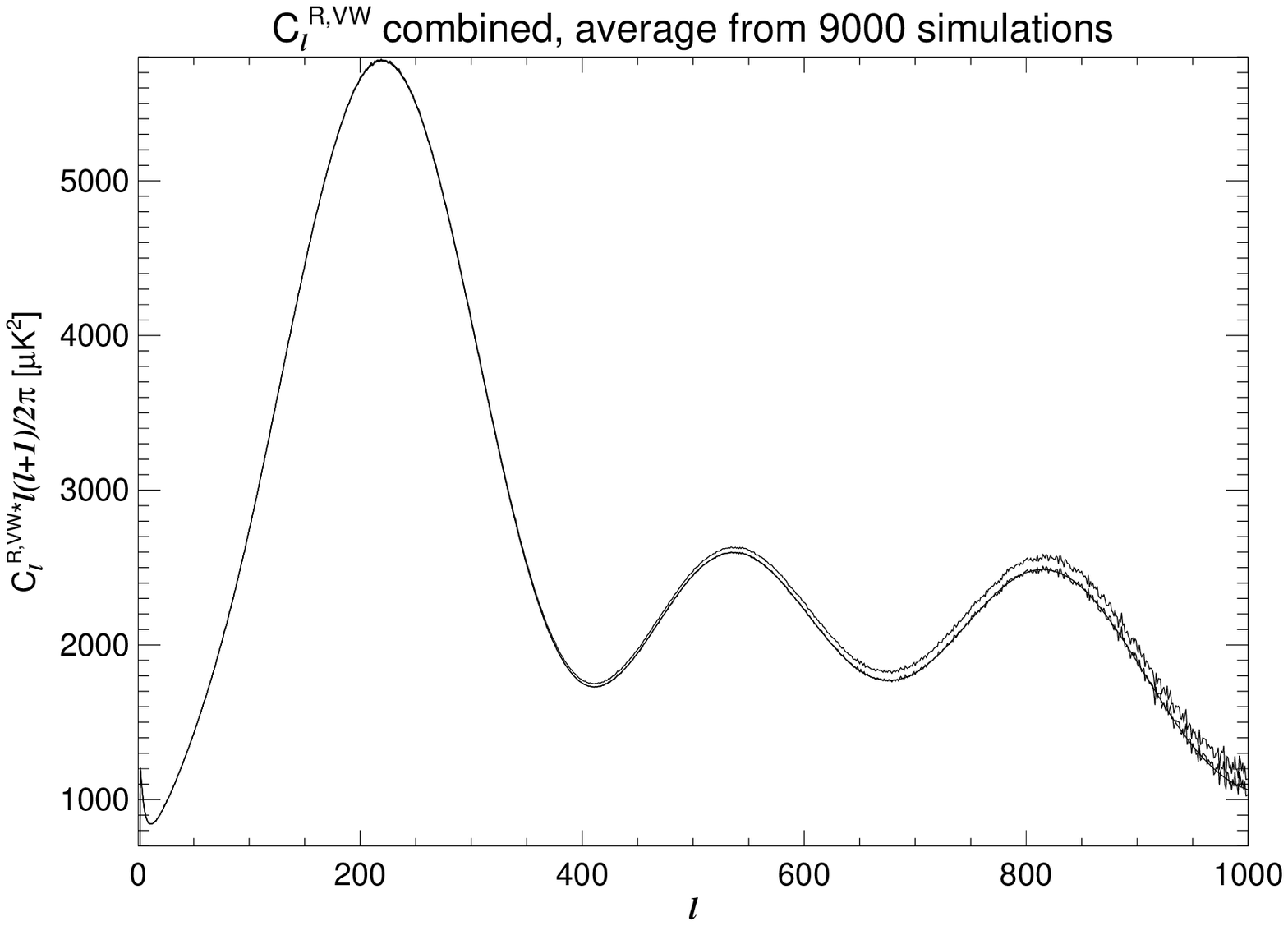}
  \includegraphics[height=9.5 cm,width= 16 cm]{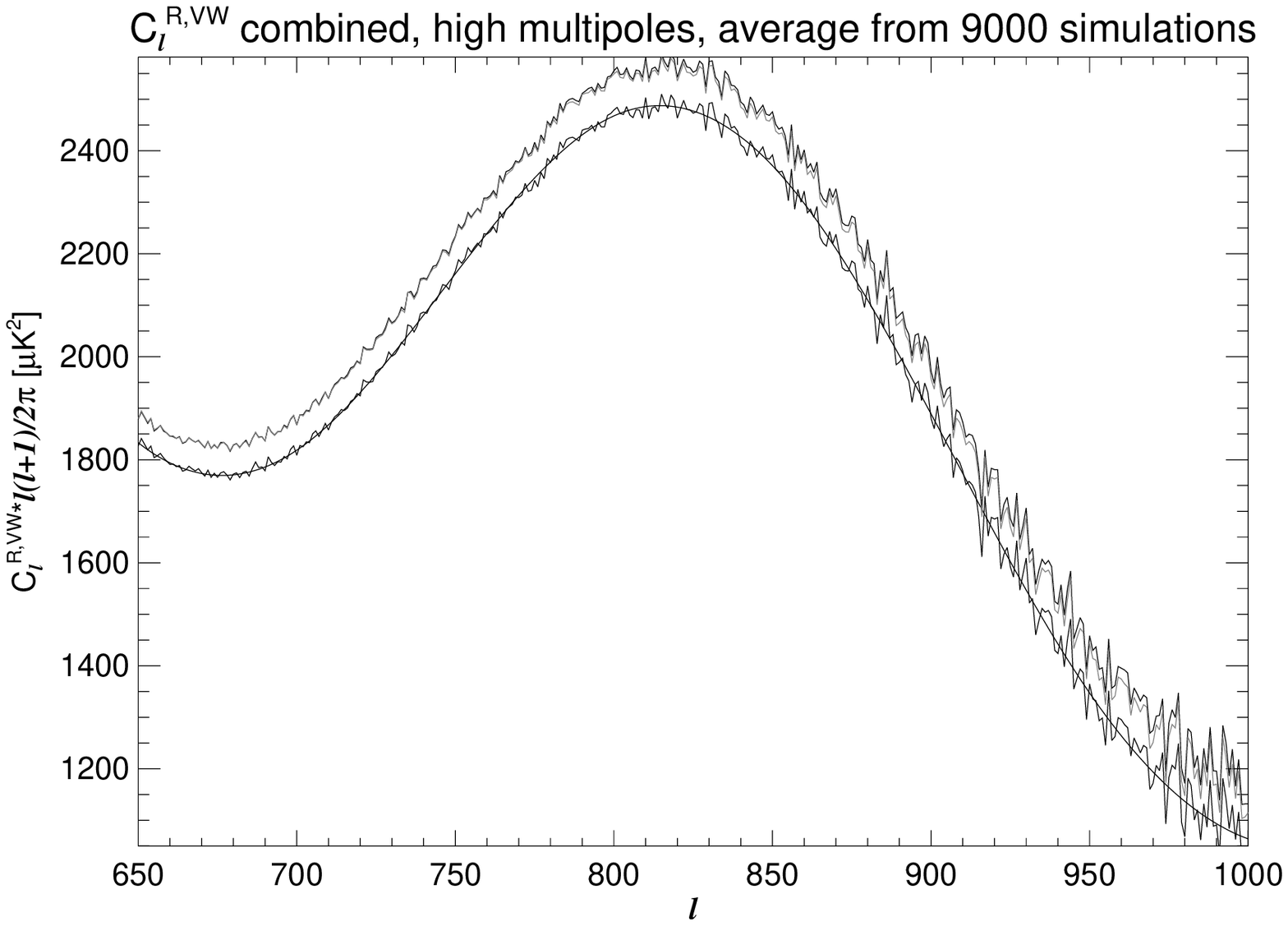}
  \caption{Average over 9000 simulations of the estimated power spectra when removing detected point sources for the combined map $M_{VW}$. In both plots the smooth curve is the input power spectrum. Top: whole $\ell$-range, before and after correcting for the contribution of unresolved sources and over-plotted the input spectrum. Bottom: Zoom to higher multipoles $\ell$, the top black curve is before correcting for the bias introduced by point source residuals, the grey curve is the power after this correction, the bottom curve is after correcting for the unresolved source contribution.}
  \label{fig:Clr_sim}
\end{figure}

\begin{figure}[h]
  \centering
  \includegraphics[scale=0.85]{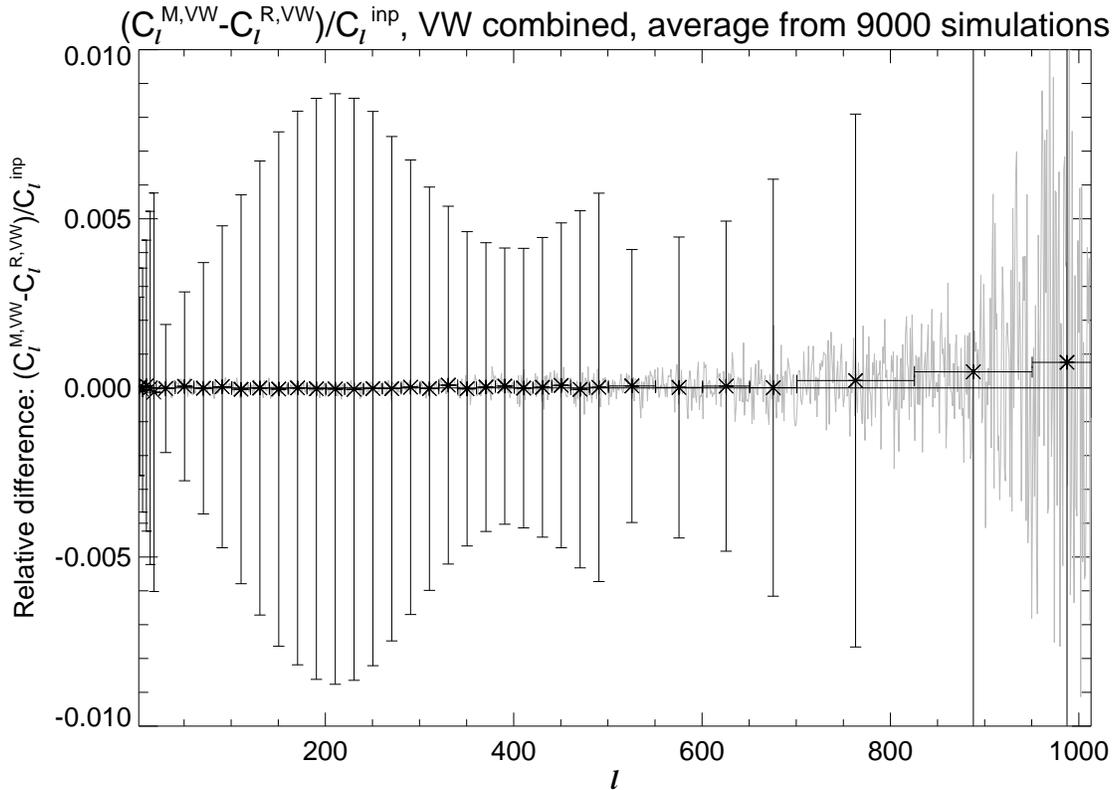}
  \caption{Relative difference, with and without binning, between masked and removed average spectra from Fig. \ref{fig:Clr_sim}. The horizontal black line represents zero. NB: for visibility, not all bins below multipole 20 are shown.}
  \label{fig:Clm_sim}
\end{figure}

In Figs. \ref{fig:Clr_sim} (removed point sources) and \ref{fig:Clm_sim} (relative difference between masked and removed) we show the average spectra obtained from the 9000 simulations from the combined map $M_{VW}$. The results are very similar for the VV, WW autospectra and VxW cross-spectrum, thus, we only show the co-added V+W results here.
In the top plot of the first figure we show the whole $\ell$-range before and after the correction of unresolved point source contribution, while in the bottom plot we show the high multipole range (600 to 1000). We also show the size of the bias introduced by the point source residuals (difference between top black curve and grey curve). As one would expect it becomes non-negligible only at high multipoles ($\ell \gtrsim 700$). 
With respect to the relative difference between the two approaches, see Fig. \ref{fig:Clm_sim}, one can remark that they are in very good agreement, having a maximal relative difference smaller than 0.1\%. 

We would like to point out that in some of the autospectra a small bias, below 1\%, is present at high multipoles, which disappears if we redo the simulations with constant spectral index. The reason for this small bias is that the method for subtracting unresolved point sources \citep{unres} assumes a constant spectral index and we found that this bias appears in the realistic case with varying spectral index. This bias is present at the same level for both source removal and source masking.

\begin{figure}[h]
  \centering
  \includegraphics[scale=0.85]{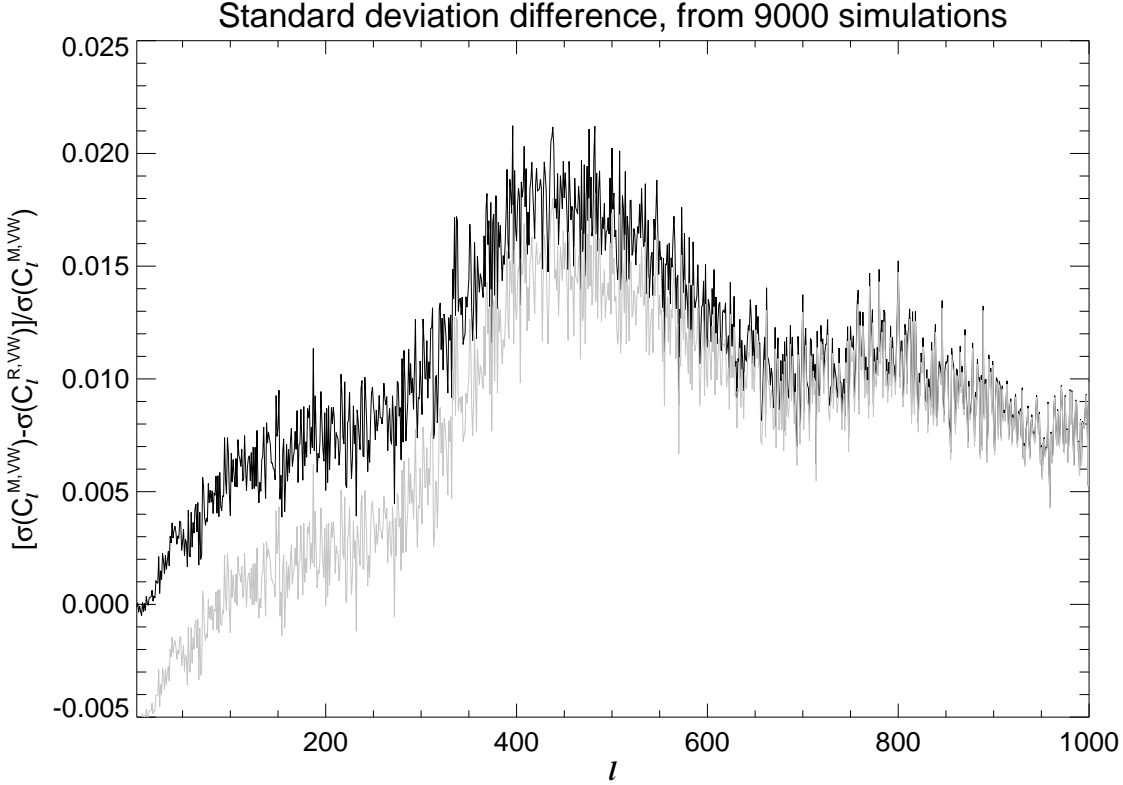}
\caption{Relative difference of the standard deviations of the estimated power spectrum of the combined map $M_{VW}$ when masking or removing, normalised to the standard deviation of the standard approach (i.e. masking, black curve). The grey curve is relative difference of the standard deviations with the cosmic variance contribution removed. Based on 9000 simulations. }
 \label{fig:stdev_sim}
\end{figure}

One of the reasons why we try two approaches (masking and removing) is that if one removes the point sources instead of masking them one uses a bigger part of the sky (and hence more information) to compute the CMB power spectrum. One would therefore expect that the standard deviation of the power spectrum found in the 9000 simulations is smaller when we remove the point sources instead of masking them. This is indeed the case as we show in Fig. \ref{fig:stdev_sim}, where we see the relative difference between the standard deviation of the power spectrum of the masking approach $\sigma(C_\ell^{M,VW})$ and the removing approach $\sigma(C_\ell^{R,VW})$ normalized to the (standard) masking approach.  We also plotted a curve where we removed the cosmic variance's contribution to the standard deviation; one can deduce from that that the difference in noise between the two methods is small but non-negligible being at maximum $\sim1.8\%$.
The increase in accuracy is expected to become substantial for the multipoles which are close in angular extension to the size of the holes (the point sources are masked with a hole of radius 0.6$^\circ\rightarrow \ell\sim 300$) and Fig. \ref{fig:stdev_sim} shows that this is indeed the case. Note that in some realistic experiments, beam asymmetries may be so large that systematic errors in the point source removal procedure may partially cancel the decrease in error bars due to increased sky fraction.

\subsection{Results: WMAP 7-year data}
\label{sec:realdata}

\subsubsection{Masking schemes}
\label{sec:msch}
As opposed to simulations, we do not only remove or mask, but (both when removing and masking) we use different kinds of masks. 
First of all we use the KQ85y7 mask (Galaxy and point sources mask) to check whether we obtain the same value of the unresolved point source contribution amplitude as the WMAP-team, obtaining:  $A_{ps}^{M,KQ85}=9.09\pm1.14$ [$10^3\mu K^2$]. This is in good agreement with the value obtained by the WMAP team: $9.0\pm0.7$ [$10^3\mu K^2$] (\cite{wmap7_power}). We point out that the uncertainty on $A_{ps}$ is now calculated the same way as by the WMAP team (see \cite{unres}). The reason that the error on the estimated $A_{ps}$ we obtain is bigger than the one of the WMAP team is due to a combination of having a different $f_{sky}(\ell)$ and having a different noise power entering the covariance matrix $\Sigma^{\alpha\beta}_{\ell\ell}$ making the estimator slightly non-optimal. These simplifications which lead to slightly larger error bars do not compromise the results of this paper.

Then we use the following masking schemes:

\begin{description}
\item[Basic mask] We only remove respectively mask the 665 detected point sources (for which we have flux densities) from the maps\footnote{the 665 point sources are detected ($5\sigma$ in the channels or $5\sigma$ in the template and $3\sigma$ in the channel) in at least Q, V or W band, not forcefully in all 3 channels, when masking we mask all 665 independently of the channel.} and use the Galaxy mask to obtain the amplitude $A_{ps}$ and the combined spectra\\ $\rightarrow$ $A_{ps}^R$ and $C_\ell^{R,\alpha}$ respectively $A_{ps}^M$ and $C_\ell^{M,\alpha}$ with $\alpha$ being a combination of channels or DAs;
\item [Extended Mask] We check for any coefficients of the needlet transform of respectively the Q-, V- and W-band maps still at $>5\sigma$ after removing or masking the point sources and extend the galactic mask with a disc of $0.6^\circ$ around these pixels.\\ $\rightarrow$ $A_{ps,ext}^R$ and $C_{\ell,ext}^{R,\alpha}$ respectively $A_{ps,ext}^M$ and $C_{\ell,ext}^{M,\alpha}$;
\item[Extended mask including KQ85] Include the KQ85y7 mask (7-year WMAP release) from the WMAP team. When masking: multiplying the extended mask by the KQ85y7 mask. When removing: only putting to zero the point sources of the KQ85y7 mask which do not coincide with any of the removed 665 sources.\\ $\rightarrow$ $A_{ps,extKQ85}^R$ and $C_{\ell,extKQ85}^{R,\alpha}$ respectively $A_{ps,extKQ85}^M$ and $C_{\ell,extKQ85}^{M,\alpha}$;
\item[All 1116] Consider all the 1116 point sources which in (\cite{newdet}) were found to be either at $5\sigma$ directly in any of the 5 WMAP frequency channels or at $5\sigma$ in internal templates and at $3\sigma$ in any of the channels. When removing: remove the above 665 sources and mask the rest.\\ $\rightarrow$ $A_{ps,1116}^R$ and $C_{\ell,1116}^{R,\alpha}$ respectively $A_{ps,1116}^M$ and $C_{\ell,1116}^{M,\alpha}$;
\item[ All 1116 including KQ85] Like before include the mask from the WMAP team.\\ $\rightarrow$ $A_{ps,1116KQ85}^R$ and $C_{\ell,1116KQ85}^{R,\alpha}$ respectively $A_{ps,1116KQ85}^M$ and $C_{\ell,1116KQ85}^{M,\alpha}$;
\item[All 2102] Consider all the 2102 different point sources which in (\cite{newdet}) were detected either at $5\sigma$ in any of the channels or at $5\sigma$ in any of the templates (without the need for a $3\sigma$ in any of the channels). Like for the 1116 approach, when removing, remove the above 665 sources and mask the rest.\\ $\rightarrow$ $A_{ps,2102}^R$ and $C_{\ell,2102}^{R,\alpha}$ respectively $A_{ps,2102}^M$ and $C_{\ell,2102}^{M,\alpha}$;
\item[All 2102 including KQ85] Like before include the mask from the WMAP team.\\ $\rightarrow$ $A_{ps,2102KQ85}^R$ and $C_{\ell,2102KQ85}^{R,\alpha}$ respectively $A_{ps,2102KQ85}^M$ and $C_{\ell,2102KQ85}^{M,\alpha}$;

\end{description}

These different masks and the pure point source maps for each DA (which were used to subtract from the temperature maps to obtain clean maps), as well as the residual corrections for the spectra obtained when removing, are available on \verb=http://folk.uio.no/frodekh/PS_catalogue/=
The only difference between the pure point source correction maps for the same frequency band is the difference in beam between the DAs. In order to use these correction maps with for instance band-averaged temperature maps, one can use one of these correction maps for one DA, deconvolve with the beam for the given DA and convolve with the desired beam. 

In Table \ref{tab:Aps} we show different values we obtain for the amplitude of the unresolved point source contribution $A_{ps}$.

\begin{table}[!hp]
\centering
\caption{Amplitude of the unresolved point source contribution, $A_{ps}$, normalised to 40.7 GHz in units $10^3\mu K^2$. The different mask names are defined in section \ref{sec:msch}.}
\vspace{0.3 cm}
\begin{tabular}{|l|c|c||c|c|c|}
\hline
                       & Value      &sky-frac\tablenotemark{a}     &     &Value      & sky-frac\tablenotemark{a} \\
\hline                                       
                       &                  &           &$A_{ps}^{M,KQ85}$     &$ 9.09 \pm   1.14$ & 78.3\%\\
\hline                                       
$A_{ps}^R$              & $10.72\pm 1.06$   & 80.6\%    & $A_{ps}^M$          & $10.05\pm    1.12$   &78.9\% \\
\hline                                       
$A_{ps,ext}^R$          &$9.19\pm   1.06$    & 80.4\%    &$A_{ps,ext}^M$       & $9.37 \pm  1.12 $  &78.7\% \\
\hline                                       
$A_{ps,extKQ85}^R$       &$7.79\pm    1.10$    & 79.4\%    &$A_{ps,extKQ85}^M$  &$ 7.95  \pm  1.16$   &77.6\% \\
\hline                                       
$A_{ps,1116}^R$         &$9.28\pm       1.10$  & 79.4\%    &$A_{ps,1116}^M$     &$8.56  \pm   1.16 $  &77.6\% \\
\hline                                       
$A_{ps,1116KQ85}^R$      &$7.10\pm     1.13$   & 78.5\%    &$A_{ps,1116KQ85}^M$  &$ 7.24  \pm   1.19$  & 76.8\%\\
\hline                                       
$A_{ps,2102}^R$         &  $8.39 \pm   1.19$   & 76.8\%    &$A_{ps,2102}^M$    & $7.55 \pm    1.26  $ &75.0\% \\
\hline                                       
$A_{ps,2102KQ85}^R$      &$6.41 \pm   1.22$ & 76.0\%    &$A_{ps,2102KQ85}^M$     &$ 6.42 \pm    1.29  $ &74.3\% \\
\hline
\end{tabular}
\label{tab:Aps}
\tablecomments{Unlike for the simulations, the error on $A_{ps}$ is obtained by the expression in \cite{unres}.}
\tablenotetext{a}{The kept sky fraction of the mask.}
\end{table}

As expected, when removing sources, the uncertainties on the estimated amplitude $A_{ps}$ are slightly smaller than when masking. This is a consequence of having a bigger sky-fraction and hence more information. In \cite{wmap5} the WMAP-team says that their estimate of $A_{ps}$ is independent of the mask used (when calculating it with Q, V and W DAs). While in our case it is very dependent on the mask. This is though no contradiction since they use always the same point-source-mask but change only the galactic cut (they use KQ75, KQ80 and KQ85), while our masks all include the same galactic mask, but with a varying number of point source holes. 

The masks with many point source holes cut a bigger part of the sky than the WMAP point source mask (in \cite{newdet} we resolved more point sources than the WMAP team) and hence the number of unresolved sources as well as their mean amplitude decreases. 

A similar behaviour of $A_{ps}$ can be seen throughout the different estimates of the unresolved point source contribution over the WMAP data releases and the corresponding number detected point sources, from 0.015 $10^3\mu K^2$ (WMAP 1-year data, number of detected point sources: 208, \cite{unres}) to 0.009 $10^3\mu K^2$ (WMAP 7-year data, number of detected point sources: 542, \cite{WMAPsrc,wmap7_power}).

\begin{figure}[h]
  \centering  
\includegraphics[width= 15 cm,height=7.3 cm]{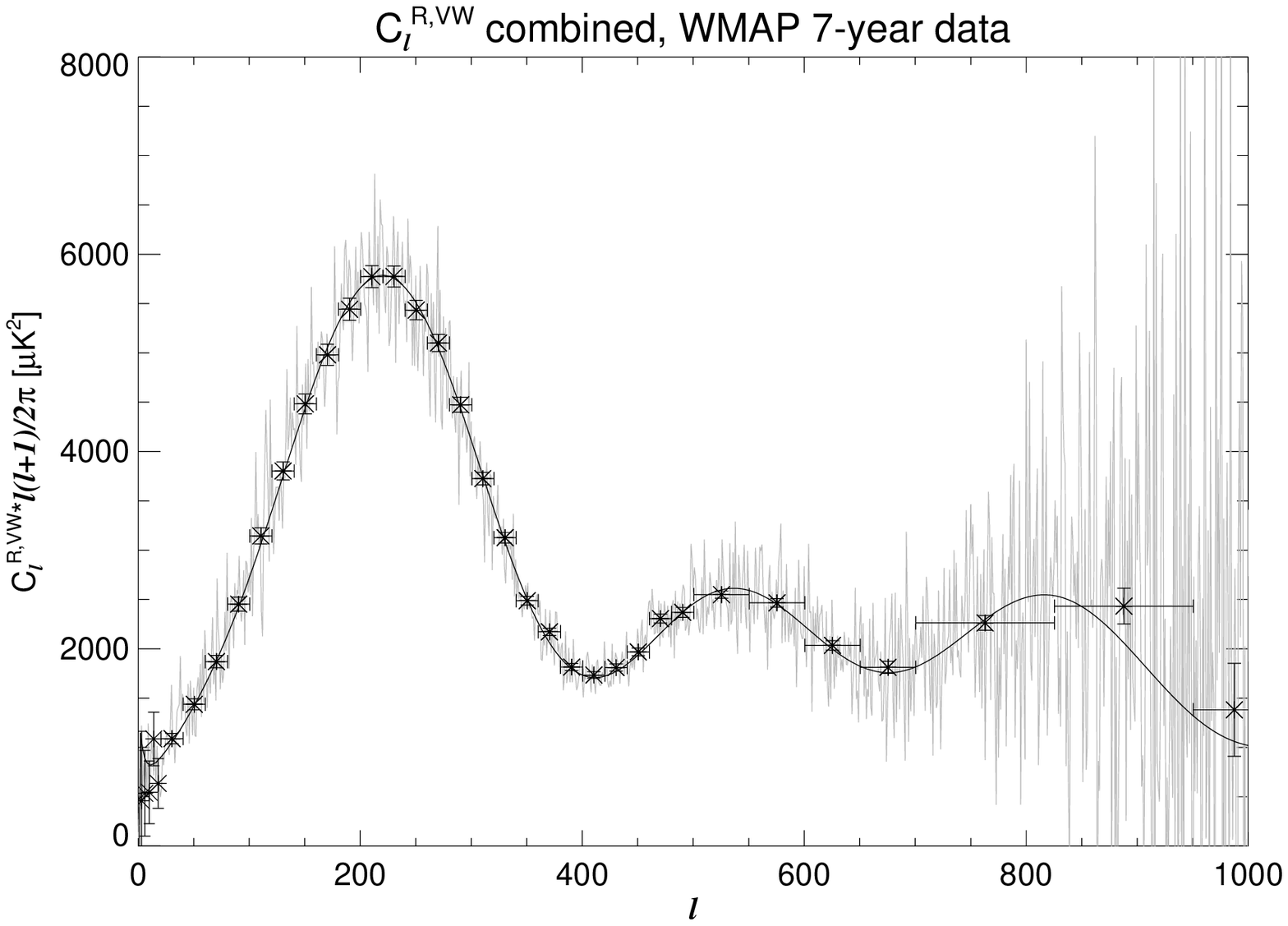}
\includegraphics[width= 15 cm,height=7.3 cm]{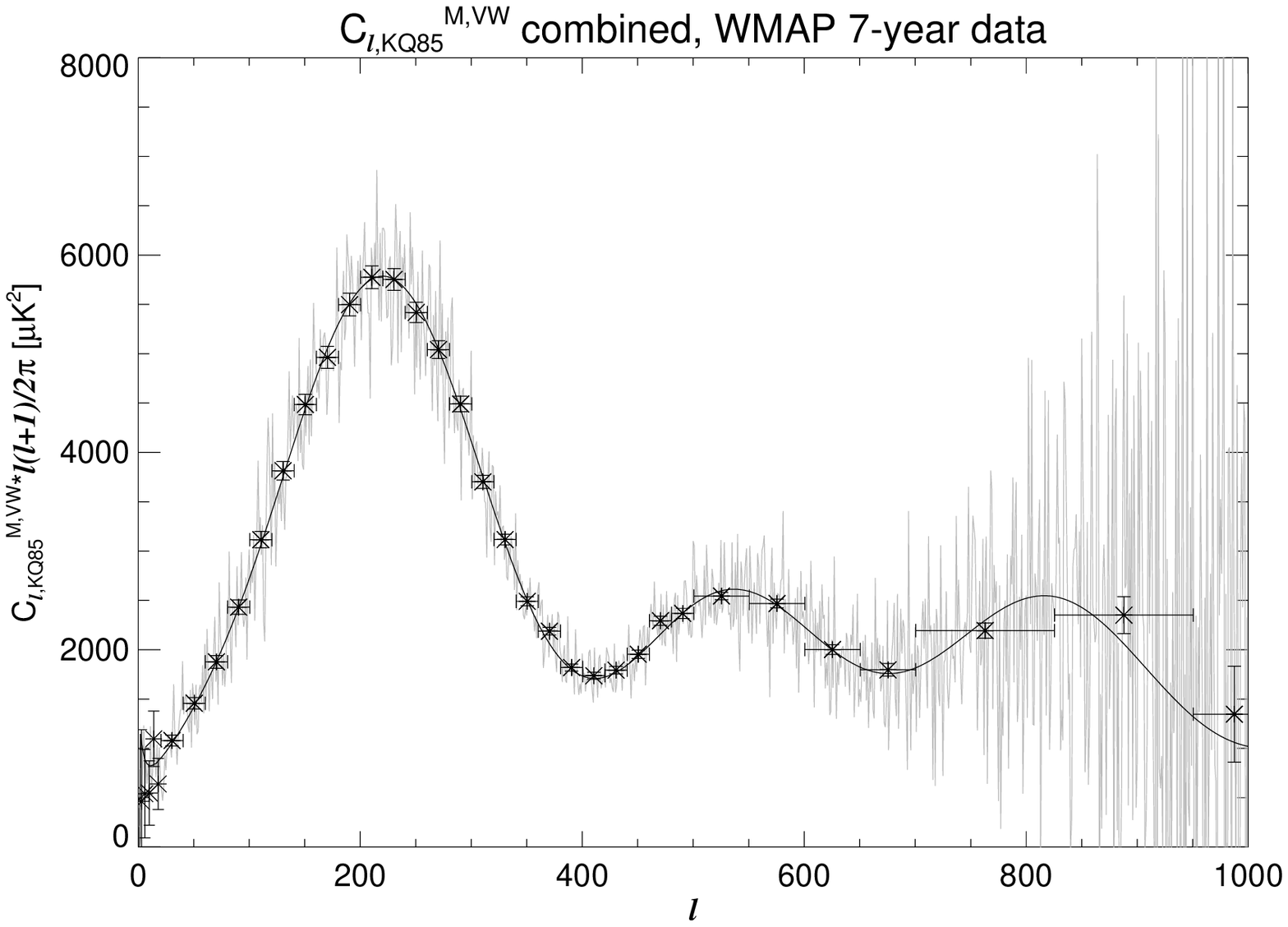}
\includegraphics[width= 15 cm,height=7.3 cm]{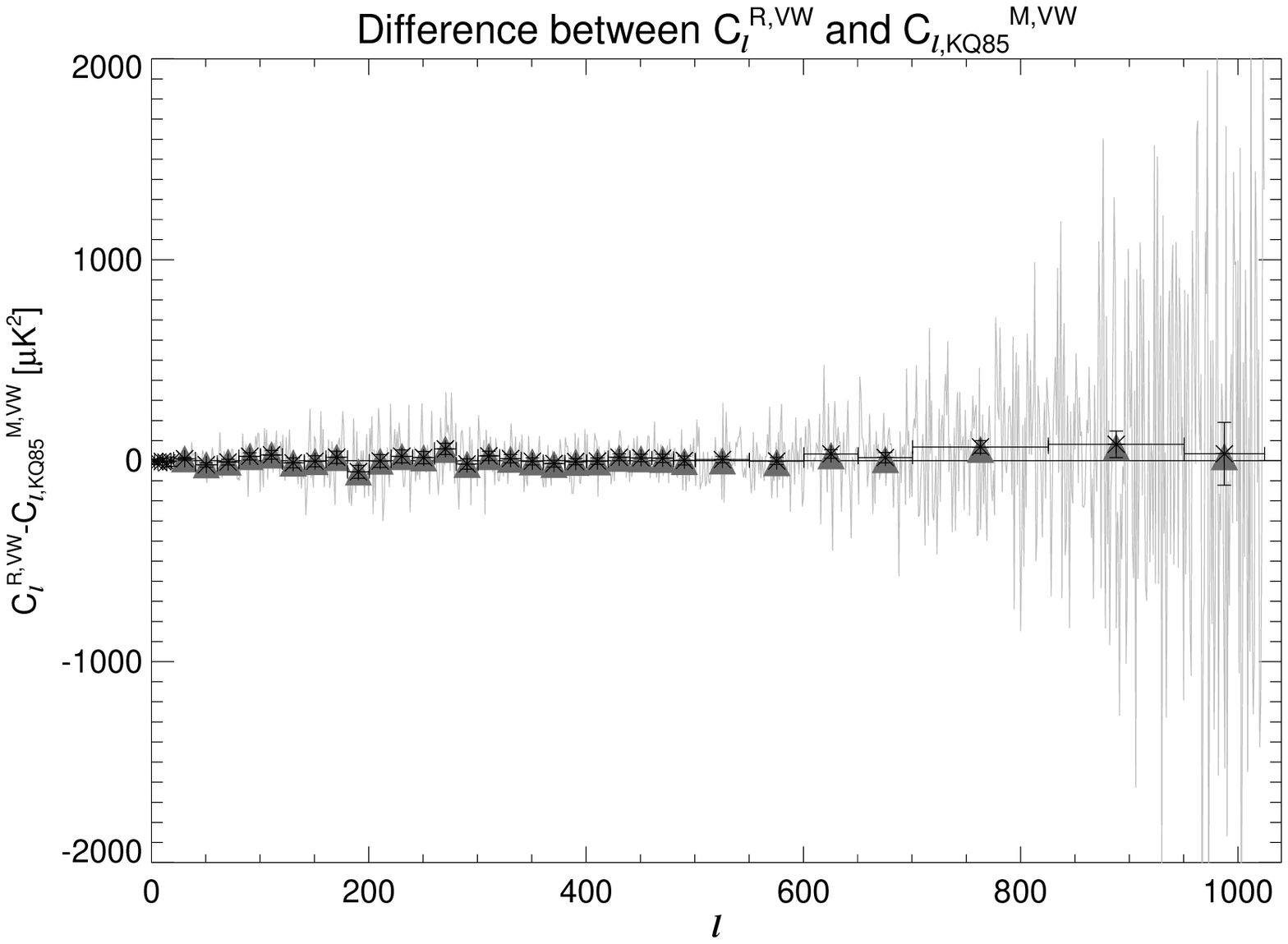}
 \caption{Example spectra, with and without binning, from the combined map $M_{VW}$ for WMAP 7-year data. Top: removing the 665 detected sources and only masking the Galaxy. Middle: using the WMAP KQ85y7 galactic and point source mask (no removal). The smooth black curve is the best-fitting WMAP-spectrum. Bottom: difference between the two spectra black line being zero (see text for details). NB: some low multipole bins are not shown.}
  \label{fig:exple}
\end{figure}

With respect to the power spectra obtained, there is no substantial difference between which mask we use or whether we remove or mask the point sources, as long as we apply the appropriate corrections for unresolved and residual sources. In the following we will show some figures for results for the CMB power spectrum obtained with the combined V+W map. The results in the VV, WW and VxW cross spectra are similar and consistent. In Fig. \ref{fig:exple} we show as an example, (1) the power spectrum obtained when removing the 665 detected point sources and masking only the Galaxy ($C_\ell^{R,VW}$) and (2) the power spectrum obtained when masking (not removing) and using only the WMAP team's KQ85y7 galactic and point source mask. We also show the difference between these spectra. These plots show well the agreement between masking and removing the detected sources when the appropriate corrections are used.
 
\noindent
In order to test the consistency of the obtained spectrum with the WMAP 7-year best-fitting spectrum (for the first two plots in Fig. \ref{fig:exple}) or with zero for the difference (third plot in same figure), we implemented a $\chi^2$ test. We compare the $\chi^2$ value of the data to the ones of 9000 simulations. It turns out that 23\% (for the removing-approach spectrum $C_\ell^{R,VW}$), 9\% (for the masked spectrum with the KQ85y7 mask $C_{\ell,KQ85}^{M,VW}$) and respectively 42\% (difference of the two spectra) of the simulations have a smaller $\chi^2$ value than the one from real data. Hence the spectrum is consistent with the best-fitting WMAP 7-year spectrum for both methods as well as the two methods are consistent with each other.

In the third plot in Fig. \ref{fig:exple} one can see that the difference for the three highest bins is sligthly bigger than zero. This is consistent with fluctuations in the determination of the amplitude of the unresolved sources $A_{ps}$. This is shown with the dark grey triangles, the values of which were obtained by changing by $1\sigma$ the value of $A_{ps}$ for the masked spectrum.

\begin{figure}[!h]
  \centering  
\includegraphics[scale=0.9]{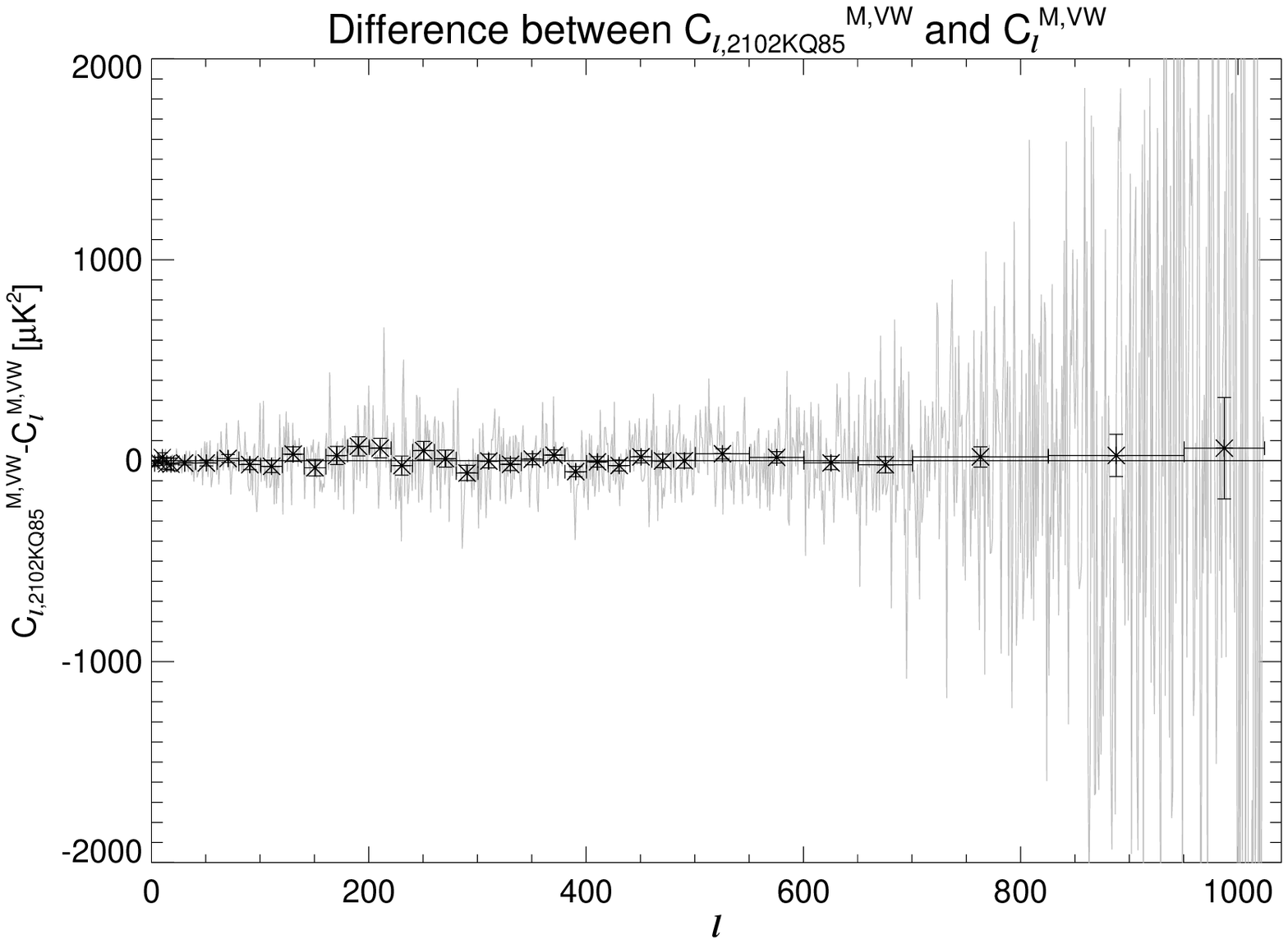}
 \caption{Difference, with and without binning, between the spectrum obtained from the combined maps of WMAP 7-year data with the biggest point source mask (2102 detected point sources and KQ85) and the one obtained with the smallest point source mask (mask of the 665 detected point sources for which we have flux estimates in the Q-, V- and W-band). The black line represents zero.  NB: some low multipole bins are not shown.}
  \label{fig:2102mQVW}
\end{figure}

In Fig. \ref{fig:2102mQVW} we show the difference between the spectra obtained with the biggest of our point source masks (Galaxy and 2102 point sources with KQ85) and the smallest point source mask (Galaxy plus 665 point sources detected in the Q-, V- and W-band) in case of only masking, with and without binning. This figure shows that the masking schemes and correction for unresolved point sources (i.e. estimation of $A_{ps}$) are well in agreement and that the weakest sources are well-corrected for using the correction for unresolved sources instead of masking. Making a $\chi^2$ test in order to check the consistency with zero of this difference, we find that 43\% of the simulations have a smaller $\chi^2$ value than the one obtained from WMAP data, hence showing well the consistency between the masking schemes.

\begin{figure}[h] 
  \centering  
\includegraphics[scale=0.85]{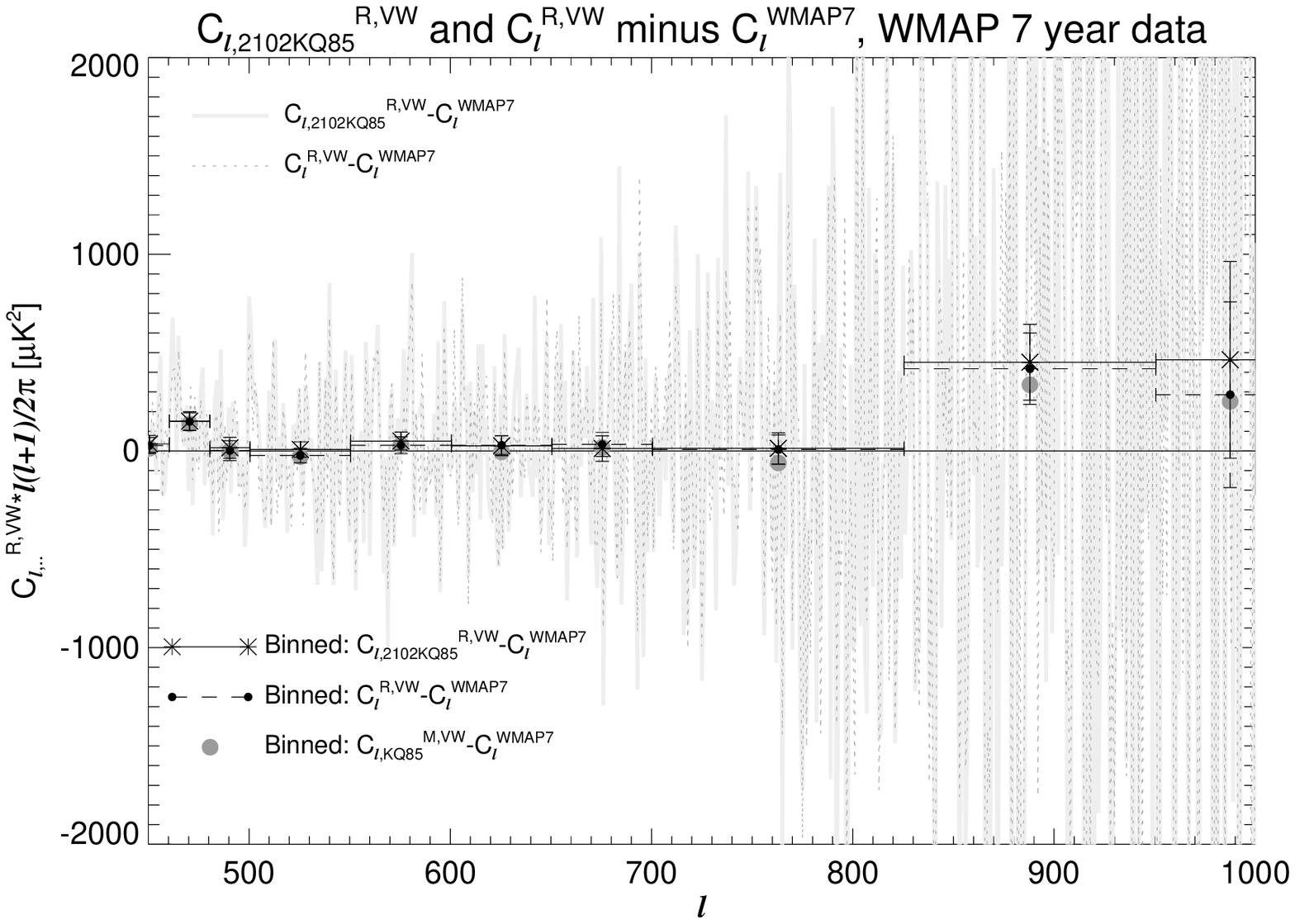}
 \caption{High multipole range, with and without binning, of the spectra $C_{\ell,2102KQ85}^{R,VW}$(grey, continuous) and $C_\ell^{R,VW}$(black, dotted) of the combined map $M_{VW}$. The best-fitting theoretical WMAP 7 spectrum has been subtracted in both cases. The grey dots represents the difference between the binned spectrum obtained with the KQ85y7 mask and the best-fitting WMAP7 spectrum. The smooth black line represents zero.}
  \label{fig:ClR2102nQVW}
\end{figure}

In Fig. \ref{fig:ClR2102nQVW} we show, with and without binning, (1) the spectrum from the maps where we remove the 665 point sources detected in Q-, V- and W-band ($C_\ell^{R,VW}$) and masking only the Galaxy and (2) the spectrum where we additionally mask the remaining of the total of 2102 point sources detected in all bands and templates ($C_{\ell,2102KQ85}^{R,VW}$). In both spectra we have subtracted the best-fitting theoretical WMAP 7 spectrum. We see that both spectra are consistent with the theoretical model and with each other.
Also the difference of the binned spectrum with the KQ85y7 mask, $C_{\ell,KQ85}^{M,VW}$, and the best-fitting theoretical WMAP 7 spectrum is shown (big grey dots).
One may expect that using a bigger sky fraction implies more information and hence less fluctuations, this figure shows that this is indeed the case (the dark grey dotted curve, $C_\ell^{R,VW}$, has clearly smaller fluctuations than the thick light gray curve, $C_{\ell,2102KQ85}^{R,VW}$).
 
We also would like to stress that when only removing the detected point sources and not using any additional masking (a part of the galactic mask) we are able to obtain consistent results (see for instance, figures \ref{fig:exple}). Hence making it possible to estimate the power spectrum $C_\ell$ without any point source masking!

\section{Skewness}
\label{sec:skew}
If there were no point sources (resolved or unresolved) or other foregrounds in the sky maps, one would expect the skewness to be zero (with the exception of tiny deviations expected from a non-zero value of the $f_{NL}$ parameter, see for instance \cite{komatsu}, where they conclude that WMAP 7-year data are consistent with $f_{NL}=0$ to 95\% CL). The presence of point sources, though, introduces a positive skewness in the maps, since they give a positive contribution to the map, which does not have a negative counterpart. In order to test our approach of removing point sources we therefore are interested in how much (if at all) the skewness changes after point source removal. We do this for 10000 simulations in both V and W channels.

In analogy to \cite{vielva04} we define the skewness $S_j$ at frequency $j$ as:
\begin{equation}
  \label{eq:skewdef}
  S_j=\frac{1}{N_{out}\cdot\sigma_j^3}\sum_k \beta_{kj}^3,
\end{equation}
 
where:
\begin{itemize}
\item $\beta_{kj}$ is the needlet coefficient at frequency $j$ and pixel $k$, multiplied by the galactic mask;
\item $N_{out}$ is the number of pixels outside the mask;
\item $\sigma_j$ is the standard deviation of the needlet coefficients, defined as:
 \[ \sigma_j=\sqrt{\frac{1}{N_{out}}\sum_k \beta_{kj}^2}.\]
\end{itemize}

The first step in our skewness analysis is to get the uncertainty (in terms of the standard deviation) of the estimated skewness $S_j$. We do this by computing the skewness for 10000 simulated maps (in both the V and W channel) with CMB and noise, but no point sources, obtaining $\sigma^{NoPS}$. We also check how much and at which frequency $j$ of the needlet skewness is introduced (in average) when there are point sources present. We therefore compute the average skewness $\langle S_j^{A}\rangle$ of 10000 maps with noise, CMB and point sources.

The next step is to see what the skewness becomes when there are only unresolved sources. We simulate 10000 maps per channel with only unresolved sources, CMB and noise but no sources above the detection limit. We then calculate the average skewness $\langle S_j^{onlyU}\rangle$ of these 10000 maps. The last step is to check whether the removal of point sources changes the skewness. In order to do this, we compute the average skewness $\langle S_j^{R}\rangle$ for 10000 maps with the noise, CMB, unresolved point sources and the detected point sources removed using best-fitting models.

\begin{figure}[h]
  \centering
\includegraphics[scale=0.85]{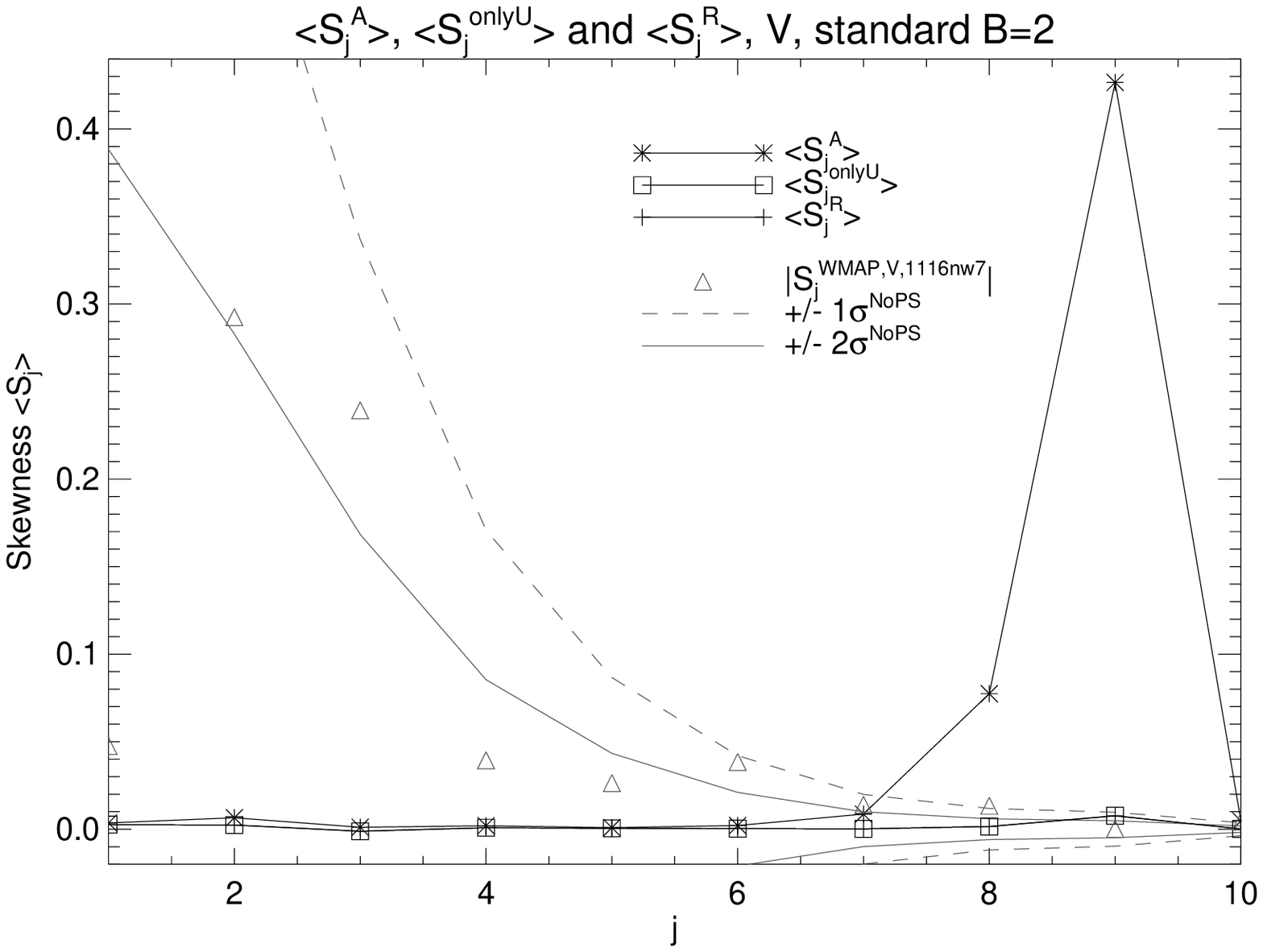}
\includegraphics[scale=0.85]{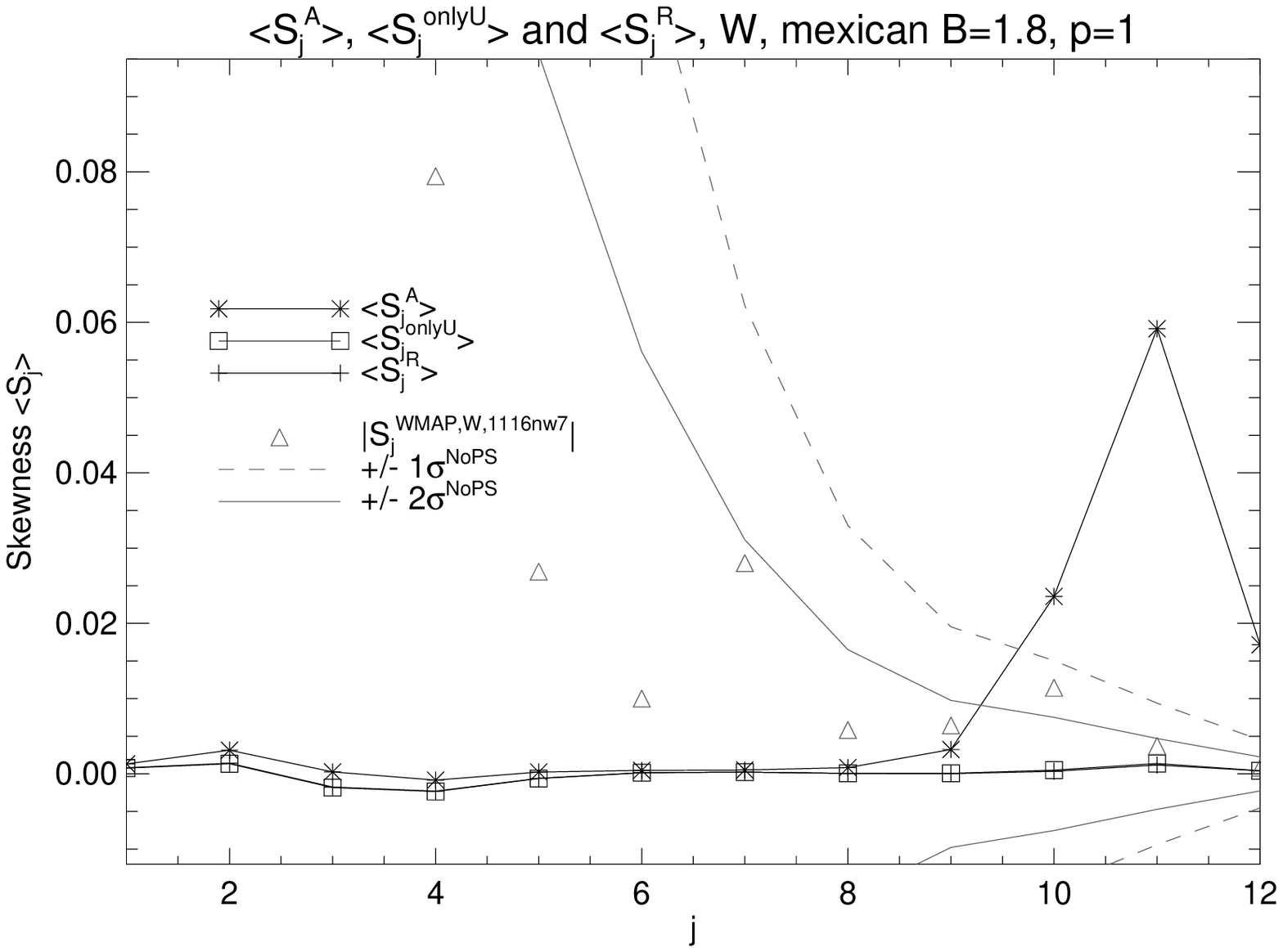}
  \caption{Average skewnesses $\langle S_j^{A}\rangle$ (black line, stars), $\langle S_j^{onlyU}\rangle$ (black line, squares) and $\langle S_j^{R}\rangle$ (black line, plus-signs). The continuous grey lines represent $\pm 1\sigma$, the dashed grey lines $\pm 2\sigma$. The grey triangles show the absolute value of the skewness with WMAP data (see text). Top: for the simulated maps of the V channel, with standard needlets and $B=2$. Bottom: for the simulated W channel maps, with mexican needlets with $B=1.8$ and $p=1$.}
  \label{fig:skew_onlyU}
\end{figure}

In figures \ref{fig:skew_onlyU} we show the average skewnesses $\langle S_j^{A}\rangle$ (black line, stars), $\langle S_j^{onlyU}\rangle$ (black line, squares)and $\langle S_j^{R}\rangle$ (black line, plus-signs) for the two needlets we use in the simulations, namely for V standard needlets with $B=2$ and for W mexican needlets with $p=1$ and $B=1.8$. These are the needlets which were used in \cite{newdet} and which were shown to be optimal for point source detection for these frequency channels. We also show the $1\sigma^{NoPS}$ (continuous grey line) and $2\sigma^{NoPS}$(dashed grey line)  intervals obtained from the standard deviation of simulated maps with no point sources. We also show an example of the absolute value of the WMAP-data skewness obtained in the case of removing the 665 detected point sources in the Q-, V- and W-band and masking the remaining of the 1116 detected point sources as well as those from the wmap KQ85y7 mask (grey triangles), as explained in \S\ref{sec:msch} (fifth entry in the masking schemes list).

As one would expect the skewness introduced by point sources (cf. $\langle S_j^{A}\rangle$) is only $\geq1\sigma$  for needlet coefficients $\beta_{j\cdot}$ with frequencies $j$ corresponding to small angular resolution. For instance, for the standard needlets (see top figure of plots \ref{fig:skew_onlyU}) there is a skewness $\geq 2\sigma$ only for frequencies $j\in[8,9]$, which correspond to $\ell$-ranges\footnote{for standard needlets the $\ell$-range seen by a needlet at frequency $j$ is given by $[B^{j-1},B^{j+1}]$.} [128,512] and  [256,1024], which in turn corresponds to angular extensions of $\sim[1.4,0.35]$ and $\sim[0.7,0.17]$ degrees. This shows why there is no visible skewness at frequencies $j$ smaller than 8, since it would imply that the point source is visible over an angular extension bigger than approximately 1 degree. 

There is no substantial difference of the skewness of the maps where we simulated and then removed the detected point sources ($\langle S_j^{R}\rangle$) and the maps where we only simulated unresolved point sources ($\langle S_j^{onlyU}\rangle$). It follows directly that point source removal does not introduce any skewness. If this were not the case, it would mean that our estimate of the amplitude of the point sources is biased. This skewness analysis hence consolidates the point source removal approach.

\section{Conclusion}
\label{sec:concl}

In this work we have studied (1) whether masking all the additional WMAP point sources reported in \cite{newdet} changes the WMAP estimate of the power spectrum compared to the spectrum obtained using the WMAP point source mask with much less point sources masked and (2) whether removing best-fitting source models instead of masking the detected point sources yields consistent results.

We compared the masking and removing approaches by making realistic simulations of 9000 maps with CMB, noise and point sources in each of the V and W channels (see \S \ref{sec:create_sims} (creating simulations) and \S\ref{sec:sim_res} (analysis of simulations)). In Figs. \ref{fig:Clr_sim} (removed point sources) and \ref{fig:Clm_sim} (difference between masked and removed point sources) we show that when estimating the $C_\ell$ both approaches give consistent results. When computing the standard deviation (see Fig. \ref{fig:stdev_sim}) we see that, as expected since there is more information, the standard deviation of the estimated power spectrum when removing the point sources is smaller than when they are masked. We remark though that when removing point sources one has to account for the bias introduced by the uncertainty on estimating the point source position and amplitude.

When considering WMAP 7-year data (\S\ref{sec:realdata}) we also compare the two approaches estimating the CMB power spectrum. But additionally we also use different masks removing different fractions of the 2102 point sources detected in \cite{newdet}. After removing respectively masking with the different masks (and correcting for the bias introduced by residual sources in case of removing), the two approaches and different masks give rise to different values of the amplitude $A_{ps}$ of the unresolved point source contribution, the values are summarized in table \ref{tab:Aps}. We remark that when using the mask of all 2102 point sources  detected at $5\sigma$ in either channels or templates, our estimate of $A_{ps}$ is substantially smaller than the estimate from the WMAP-team: $6.41 \pm 1.22$ (removing and masking) and $6.42 \pm 1.29$ (only masking) versus $9.0 \pm 0.7$ from the WMAP-team with the KQ85y7 mask (all amplitudes $A_{ps}$ in units of $[10^3\mu K^2]$). Hence this new mask presents a big improvement, with less contamination by unresolved point sources.

We found that the power spectrum obtained when removing the 665 detected sources for which we have flux density estimates and using only the galactic mask (no point source mask), is fully consistent with the power spectrum obtained when masking all the 2102 detected sources as well as those from the KQ85y7 point source mask. The consistency amongst different masking schemes and methods to obtain the power spectrum was tested with a $\chi^2$ test, yielding values of $\chi^2$ for real data, well between $\chi^2$ values obtained from corresponding simulations (see \S\ref{sec:realdata} for numbers). This shows (1) that it suffices to take into account only the sources detected by the WMAP team and correct for the remaining sources through the unresolved source correction and (2) that removing the best-fitting source models works as well as masking them. The advantage of removing is that for some CMB analysis methods (especially wavelet/filter based methods), the point source holes reduce significantly the sky fraction which can be used on the filtered maps. Note however, that in the case of point source removal, the appropriate correction for residual sources has to be applied. The removing approach is also verified with respect to skewness, where we showed that the fact of removing point sources does not introduce any form of skewness, showing that the removal of point sources is an unbiased approach.

The different masks, the different pure point source maps for each DA (which can be used to remove sources in temperature maps to obtain clean maps) and the residual power corrections are available on: \verb=http://folk.uio.no/frodekh/PS_catalogue/=

\acknowledgments 

SSc would like to thank S.K. N\ae ss for useful discussions. FKH acknowledges an OYI grant from the Norwegian Research Council. Super computers from NOTUR (The Norwegian metacenter for computational science) have been used in this work. We acknowledge the use of the HEALPix software package (\cite{healpix}) and the Legacy Archive for Microwave Background Data Analysis (LAMBDA) to retrieve the WMAP data set. This research has made use of data obtained from the High Energy Astrophysics Science Archive Research Center (HEASARC), provided by NASA's Goddard Space Flight Center and of the NASA/IPAC Extragalactic Database (NED) which is operated by the Jet Propulsion Laboratory, California Institute of Technology, under contract with the National Aeronautics and Space Administration.

\end{document}